\documentclass[aps, pra, reprint, lengthcheck, footinbib, showpacs,  citeautoscript]{revtex4-2}

\usepackage[T1]{fontenc}

\usepackage{multirow}

\usepackage{amsmath}

\usepackage{amsmath}

\usepackage{braket}
\usepackage{graphicx}

\usepackage[colorlinks, urlcolor=blue, citecolor=blue, linkcolor=blue, pdfstartview=FitH]{hyperref}
\usepackage[all]{hypcap}

\usepackage{dcolumn}
\usepackage{bm}
\usepackage{epstopdf}

\usepackage{xcolor} 
\usepackage[normalem]{ulem}
\definecolor{color}{rgb}{0.11,0.45,0.02}

\begin{document}

\title{Hole spin precession and dephasing induced by nuclear hyperfine fields in CsPbBr$_3$ and CsPb(Cl,Br)$_3$ nanocrystals in a glass matrix}

\begin{abstract}

The coherent spin dynamics of holes are investigated for CsPbBr$_3$ and CsPb(Cl,Br)$_3$ perovskite nanocrystals in a glass matrix using the time-resolved Faraday rotation/ellipticity techniques. In an external magnetic field, pronounced Larmor spin precession of the hole spins is detected across a wide temperature range from 5 to 300~K. The hole Land\'e $g$-factor varies in the range of $0.8-1.5$, in which it increases with increasing high energy shift of the exciton due to enhanced confinement in small nanocrystals. The hole spin dephasing time decreases from 1 ns to 50 ps in this temperature range. Nuclear spin fluctuations have a pronounced impact on the hole spin dynamics. The hyperfine interaction of the holes with nuclear spins modifies their spin polarization decay and induces their spin precession in zero external magnetic field. The results can be well described by the model developed in Ref.~\cite{Merkulov2002}, from which the hyperfine interaction energy of a hole spin with the nuclear spin fluctuation  in range of $2-5$~$\mu$eV is evaluated. 
 
\end{abstract}

\author{Sergey~R.~Meliakov$^{1}$, Vasilii~V.~Belykh$^{2}$, Evgeny~A.~Zhukov$^{2,1}$,  Elena~V.~Kolobkova$^{3,4}$, Maria~S.~Kuznetsova$^5$, Manfred~Bayer$^{2}$, Dmitri~R.~Yakovlev$^{2,1}$}

\affiliation{$^{1}$P.N. Lebedev Physical Institute of the Russian Academy of Sciences, 119991 Moscow, Russia}
\affiliation{$^{2}$Experimentelle Physik 2, Technische Universit\"at Dortmund, 44227 Dortmund, Germany}
\affiliation{$^{3}$ITMO University, 199034 St. Petersburg, Russia}
\affiliation{$^{4}$St. Petersburg State Institute of Technology, 190013 St. Petersburg, Russia}
\affiliation{$^{5}$Spin Optics Laboratory, St. Petersburg State University, 198504 St. Petersburg, Russia}

\date{\today}

\maketitle

\section{Introduction}

The semiconductors based on the lead halide perovskite compounds have attracted the vast attention of physicists and chemists in the last decade. The interest has been inspired by the discovery of their large photovoltaic response, their bright emission tunable over the whole visible spectral range and the relatively low-demand fabrication technology~\cite{Jena2019, Vinattieri2021_book, Vardeny2022_book}. This has stimulated research in many different directions, suggesting multiple applications of perovskite semiconductors and of their nanostructures.

The lead halide perovskites have the chemical formula $A$Pb$X_3$, where the cation $A$ can be cesium (Cs), methylammonium (MA), or formamidinium (FA), and the anion $X$ can be Cl, Br, or I. The materials with Cs-cation are fully-inorganic, while with MA or FA they are hybrid organic-inorganic. The lead ions dominate the electronic and spin properties of the states forming the band gap~\cite{Amat2014,Xiao2019,kirstein2022nc,kirstein2022am}. Synthesis in solution is the most popular technique to grow single crystals, polycrystalline films, and nanocrystals (NCs). Both hybrid organic-inorganic and fully-inorganic NCs can be grown~\cite{Kovalenko2017,Efros2021,Zhang2017,Yarita2018}, which, however, show degradation in ambient conditions, especially the NCs with organic cations. An efficient solution of this problem is isolating the NCs from their environment by synthesizing them in a glass matrix~\cite{Liu2018,Liu2018a,Li2017,Ye2019,Kolobkova2021}. This can be used only for the fully-inorganic NCs, where the process is based on cooling of the glass melt starting from temperatures of around 1000$^\circ$C. Such samples are convenient for performing optical experiments after glass polishing.     

The lead halide perovskites show attractive spin properties, which can be addressed by well-established optical and magneto-optical experimental techniques, such as time-resolved Faraday/Kerr rotation spectroscopy~\cite{belykh2019,Grigoryev2021,kirstein2022nc,kirstein2022am}, photoluminescence (PL) with polarization resolution~\cite{Giovanni2015,Nestoklon2018,kopteva2024_oo, canneson2017}, spin-flip Raman scattering~\cite{kirstein2022nc,Harkort_2D_2023}, polarization-sensitive time-resolved differential transmission~\cite{Strohmair2020}, and optically detected magnetic resonance~\cite{Belykh2022}. This allows one to revisit the spin phenomena discovered for conventional III-V and II-VI semiconductors~\cite{OOBook1984,SpinBook2017,GlazovBook2018} for different starting conditions, provided by the specifics of the perovskite band structure~\cite{kirstein2022nc,kirstein2022am}. Among these phenomena there are optical spin orientation~\cite{Nestoklon2018,kopteva2024_oo,kudlacik2024_oo_carriers}, coherent spin precession in a magnetic field~\cite{belykh2019,Grigoryev2021,kirstein2022nc}, dynamic nuclear polarization~\cite{kirstein2022am}, and a universal dependence of the $g$-factors on the band gap energy in bulk perovskites~\cite{kirstein2022nc,Kopteva_gX_2023}. Remarkably, in CsPbBr$_3$ NCs, carrier spin coherence with dephasing times up to several nanoseconds was observed~\cite{Crane2020,Lin2022,Meliakov2023NCs,Cheng2024} and the feasibility of optical spin manipulation was demonstrated~\cite{Lin2022}. For CsPbI$_3$ and CsPb(Cl,Br)$_3$ NCs in a glass matrix, coherent spin dynamics were demonstrated only at cryogenic temperatures~\cite{Belykh2022,kirstein2023_SML,nestoklon2023_nl}.

The spin-dependent phenomena are enriched by the interaction of carrier spins with nuclear spins. Cryogenic temperatures  and strong carrier localization in, e.g., quantum dots (QDs) greatly enhance the appearance of these phenomena \cite{Stepanenko2006,Chekhovich2013,Bechtold2015,Abobeih2018}. The nuclear spin system can be strongly polarized via its interaction with optically-oriented carriers. In turn, the Overhauser field of the polarized nuclei acts back on the carrier spins, causing their spin splitting and facilitating their spin dephasing. Even for an unpolarized nuclear spin system, the presence of spin fluctuations may control the spin dynamics of localized carriers. This problem was considered theoretically in 2002 by Merkulov, Efros and Rosen~\cite{Merkulov2002}. Its most straightforward manifestation in form of a half-period precession of the oriented spin at zero external magnetic field was reported for electron spins in GaAs-based QDs~\cite{Bechtold2015} and Ce$^{3+}$ ions in YAG crystals~\cite{Liang2017}. In III-V and II-VI semiconductors, the electrons in the conduction band with s-type wave function interact much more strongly with the nuclei than the holes in the valence band with p-type wave function. The situation is reversed in the lead halide perovskites, where the valence band is strongly contributed by s-orbitals of the Pb-ions and the conduction band is formed by p-orbitals of the Pb-ions. As a result, the hole-nuclei interaction is about an order of magnitude stronger than the electron-nuclei interaction~\cite{kirstein2022am}. 

The role of the nuclear spin fluctuations in the spin dephasing of localized electrons and holes was studied in a bulk FA$_{0.9}$Cs$_{0.1}$PbI$_{2.8}$Br$_{0.2}$ perovskite crystal by means of the Hanle and polarization recovery effects~\cite{kudlacik2024_oo_carriers}. For this crystal, optically-detected magnetic resonance was demonstrated by the combination of time-resolved Kerr rotation with additional radiofrequency excitation~\cite{kirstein2022am}. Dynamic nuclear polarization was demonstrated in crystals of FA$_{0.9}$Cs$_{0.1}$PbI$_{2.8}$Br$_{0.2}$~\cite{kirstein2022am}, MAPbI$_3$~\cite{kirstein2022mapi}, FAPbBr$_3$~\cite{Kirstein2023DNSS}, and CsPbBr$_3$~\cite{belykh2019}. In FAPbBr$_3$ crystals, the suppression of the nuclear spin fluctuations via creation of a squeezed dark nuclear spin state was demonstrated~\cite{Kirstein2023DNSS}. The state formation by optical excitation is governed by quantum correlations and entanglement between the nuclear spins. For CsPb(Cl,Br)$_3$ NCs in glass, the spin mode-locking effect for holes was shown to be stabilized by the nuclear frequency focusing effect, where the Larmor precession frequencies of hole spins in each NC in an ensemble were adjusted by the respective nuclear polarization to achieve synchronization with the pump laser frequency~\cite{kirstein2023_SML}.

In this paper, time-resolved Faraday rotation/ellipticity is used to study the coherent spin dynamics of carriers in CsPbBr$_3$ NCs in a glass matrix across a wide temperature range from 5 to 300~K in order to measure the temperature dependences of the $g$-factors and spin relaxation times. From the $g$-factor spectral and temperature dependences we identify the hole to be the origin of the Larmor spin precession signals in finite magnetic fields. At zero magnetic field, hole spin precession in the random nuclear exchange fields is detected and the parameters of the nuclear spin fluctuations are evaluated. This effect is also found for CsPb(Cl,Br)$_3$ NCs in a glass matrix.  

\section{Experimentals}

The CsPbBr$_3$ and CsPb(Cl,Br)$_3$ NCs were synthesized in a matrix of fluorophosphate glass using the melt-quench technique. Technological details and information on the sample characterization can be found in Refs.~\onlinecite{Kolobkova2021,Kulebyakina2024,kirstein2023_SML}. In this paper we study two samples with CsPbBr$_3$ NCs. One sample contains NCs with average size of 9~nm (sample \#1, EK1) and a broad size distribution in the range of about $7-11$~nm. Another sample contains 16~nm NCs (sample \#2, EK103) with a narrow size distribution. The third sample hosts CsPb(Cl$_{0.56}$Br$_{0.44}$)$_3$ NCs with average size of 8~nm (sample \#3, EK14). Detailed information on photoluminescence of these samples, including the recombination dynamics in the temperature range $5-270$~K is given in Ref.~\onlinecite{Kulebyakina2024}. The coherent spin dynamics of holes at cryogenic temperatures were reported for sample \#3 in Refs.~\onlinecite{kirstein2023_SML,Belykh2022}. 

The photoluminescence (PL) spectra are measured for excitation with a continuous-wave semiconductor laser with the photon energy of 3.06~eV (wavelength of 405~nm). The emission is recorded with an 0.5-meter spectrometer equipped with a liquid-nitrogen-cooled charge-coupled-device camera with a silicon matrix. The samples were placed in a helium-flow optical cryostat at the temperature of $T=5$~K.  

To study the coherent spin dynamics of carriers we use a time-resolved pump-probe technique with detection of the Faraday rotation (TRFR) or Faraday ellipticity (TRFE)~\cite{Yakovlev_Ch6}. Spin-oriented electrons and holes are generated by circularly polarized pump pulses. The used laser system (Light Conversion) generates 1.5-ps pulses with a spectral width of about 1~meV at a repetition rate of 25~kHz (repetition period 40~${\mu}$s). The laser photon energy is tuned in the spectral range of $2.35 - 2.80$~eV in order to provide resonant excitation of NCs of specific sizes. The laser beam is split into the pump and probe beams with coinciding photon energies. The time delay between the pump and probe pulses is adjusted by a mechanical delay line. The pump beam is modulated with an electro-optical modulator between ${\sigma}^+$ and ${\sigma}^-$ circular polarization at the frequency of 26~kHz. The probe beam is linearly polarized. The Faraday rotation angle (or Faraday ellipticity) of the probe beam, which is proportional to the carrier spin polarization, is measured as function of the delay between the pump and probe pulses using a balanced photodetector, which is connected to a lock-in amplifier synchronized with the modulator. Both pump and probe beams have powers of 0.5~mW and spot sizes on the sample of about 100$~\mu$m. For the time-resolved measurements the samples are placed in a helium-flow optical cryostat and the temperature is varied in the range of $5-300$~K. Magnetic fields up to 500~mT are applied perpendicular to the laser wave vector (Voigt geometry, $\textbf{B} \perp \textbf{k}$) using an electromagnet.

\section{Results and discussion}

\begin{figure}[hbt!]
\includegraphics[width=0.9\columnwidth]{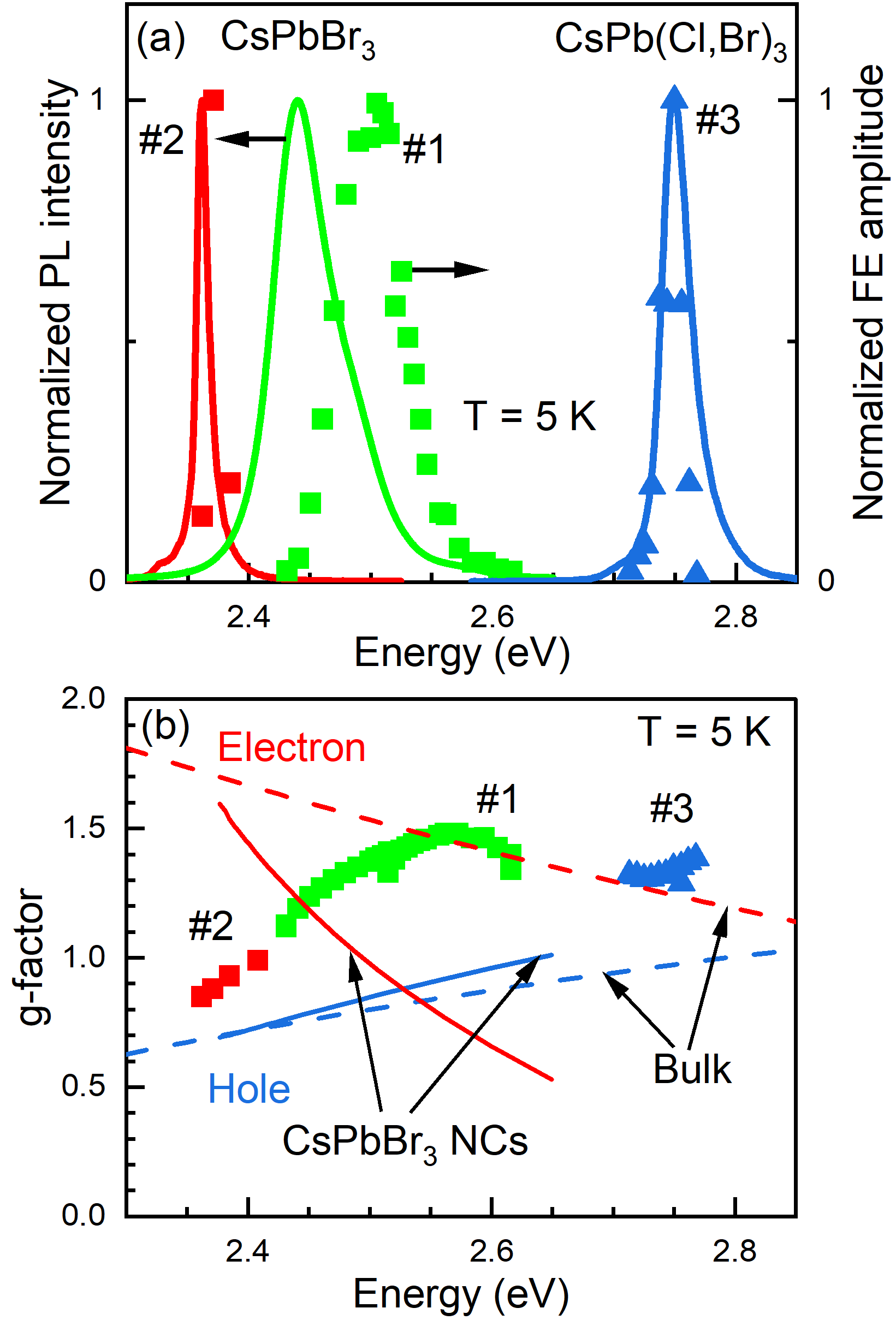} 
\caption{Spectral dependences of the spin parameters for the samples \#1 (green), \#2 (red), and \#3 (blue), measured at $T=5$~K. 
(a) Normalized PL spectra (solid lines) and spectral dependences of the Faraday ellipticity amplitude (symbols).
(b) Spectral dependence of the measured $g$-factor (symbols). Dashed lines show the universal dependences of the electron (red) and hole (blue) $g$-factors on the band gap energy for bulk crystals~\cite{kirstein2022nc}. Solid lines show the calculated dependences of the electron (red) and hole (blue) $g$-factors on the effective band gap in CsPbBr$_3$ NCs~\cite{nestoklon2023_nl}.
}
\label{fig:Spectral}
\end{figure}

Figure~\ref{fig:Spectral}a shows PL spectra for the studied samples, measured at $T=5$~K. The emission lines have maxima at the energies of 2.441~eV, 2.361~eV, and 2.749~eV for samples \#1, \#2, and \#3, respectively. The width of the PL spectra is given by the spread of NCs sizes, which is largest for sample \#1. The different spectral positions of the PL peaks for the CsPbBr$_3$ NCs is due to the different quantum confinement energy of carriers. The spectral shift is larger for sample \#1 with smaller NCs compared to  sample \#2. The spectral position of the PL peak for sample \#3 is mainly determined by inclusion of Cl in the CsPb(Cl,Br)$_3$ NCs and partly by carrier confinement.   

\subsection{Coherent spin dynamics of holes at $T=5$~K in CsPbBr$_3$ NCs}

\begin{figure*}[hbt!]
\centering
\includegraphics[width=1.7\columnwidth]{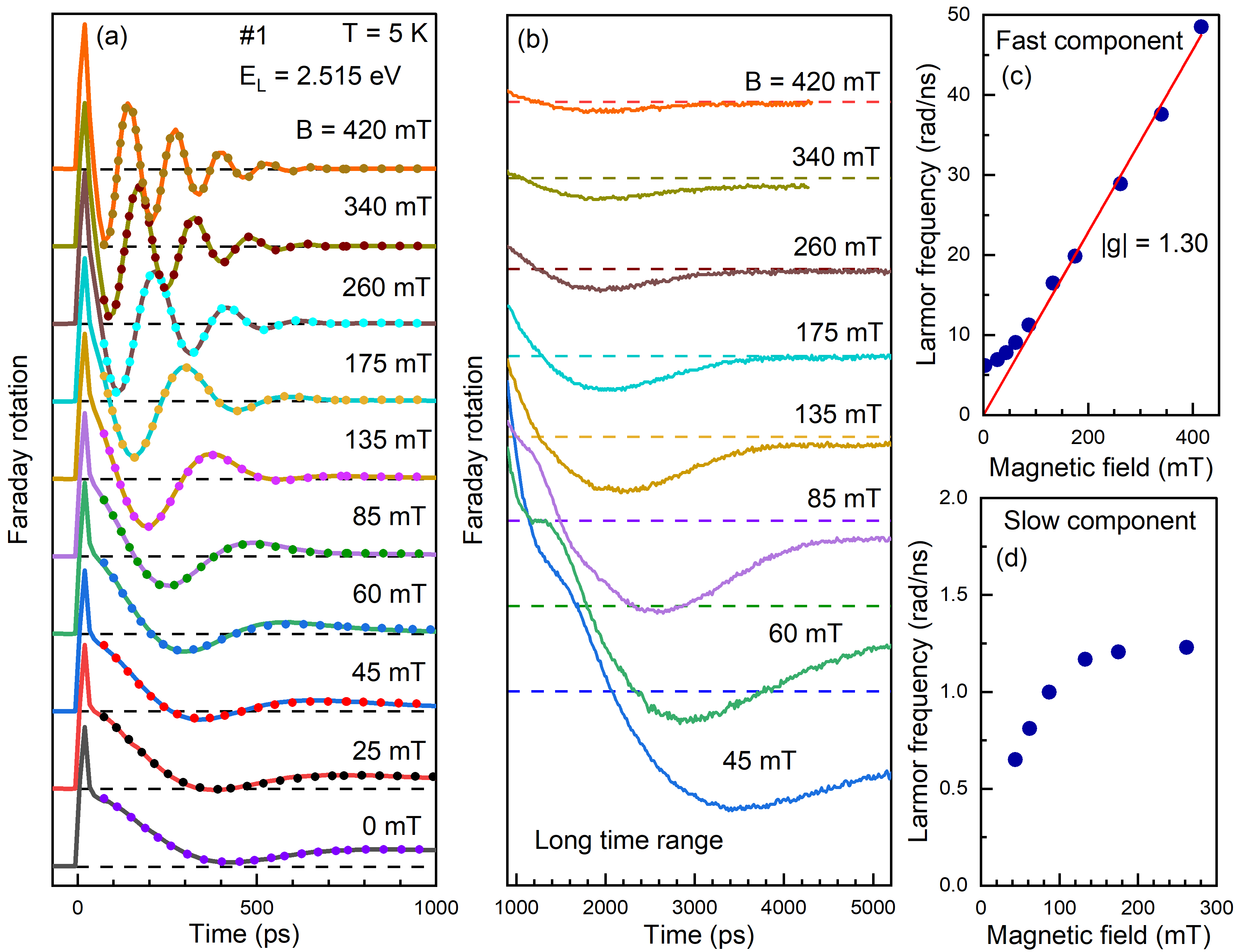} 
\caption{Magnetic field dependence of the spin dynamics in CsPbBr$_3$ NCs (sample~\#1). $E_L= 2.515$~eV, $T=5$~K.
(a) TRFR traces in magnetic field varied from 0 to 420~mT. Fits to the data using the Merkulov-Efros-Rosen model with an exponential decay are shown by the dots.
(b) TRFR traces using a magnified vertical scale at later time delays compared to (a). A component with low oscillation frequency is observed.
(c) Dependence of the Larmor precession frequency from the fast oscillating component on magnetic field. The red line shows a fit with Eq.~\eqref{eq:LarmFreq}, giving $|g|=1.30$.
(d) Dependence of the Larmor precession frequency from the slowly oscillating component on magnetic field.
}
\label{fig:Magnetic}
\end{figure*}

We use time-resolved Faraday rotation to study the coherent spin dynamics of carriers in CsPbBr$_3$ NCs. The dynamics show the most pronounced features at the low temperature of $T=5$~K. The representative spin dynamics measured on sample~\#1 in different magnetic fields up to 420~mT are shown in Fig.~\ref{fig:Magnetic}(a). The laser photon energy is set to $E_\text{L}=2.515$~eV, corresponding to the maximum of the FR signal. The dynamics show the expected behavior for an ensemble of carrier spins precessing about a perpendicular magnetic field with the Larmor frequency, $\omega_\text{L}$, which spin polarization decays with the spin dephasing time, $T_2^*$~\cite{Yakovlev_Ch6}. With increasing magnetic field the Larmor precession frequency grows and the spin dephasing time shortens.  

These spin dynamics are commonly fitted with the equation:
\begin{equation}
A_{\rm FR}(t) \propto S_0  \cos(\omega_\text{L}t) \exp(-t/{T_2^*})  ,
\label{eq:VoightTRFR}
\end{equation}
where $S_0$ is the initial spin polarization of the photogenerated carriers along the optical axis and $t$ is the time delay between the pump and probe pulses. The fit allows us to determine the Larmor precession frequencies, which are shown in Fig.~\ref{fig:Magnetic}(c) as function of the magnetic field.  At $B>80$~mT the dependence $\omega_\text{L}(B)$ is linear corresponding to 
\begin{equation}
\omega_\text{L}(B) = |g|{{\mu}_\text{B}}B/{\hbar} ,
\label{eq:LarmFreq}
\end{equation}
where ${\mu}_\text{B}$ is the Bohr magneton and $g$ is the carrier Land\'e $g$-factor. The slope of the linear dependence in this range, shown by the red line, gives $|g|=1.30$. The sign of the $g$-factor cannot be directly determined from TRFR signals, but it is positive for both electrons and holes in bulk CsPbBr$_3$~\cite{kirstein2022nc} and is theoretically predicted to maintain this sign for not too small CsPbBr$_3$ NCs~\cite{nestoklon2023_nl}. Therefore, we take it as positive in what follows and list our arguments for assigning the observed spin signals to the hole spin precession below. 

An interesting and rather unusual result is the observation of oscillations in the spin dynamics even at zero magnetic field, see the bottom graph in Fig.~\ref{fig:Magnetic}(a). One also sees in Fig.~\ref{fig:Magnetic}(c) that with decreasing magnetic field the Larmor precession frequency does not go to zero, but saturates at a value of about 6~rad/ns, which corresponds to the Zeeman splitting of 4~$\mu$eV. We note, that zero-field spin precession with a much higher frequency of about 2~rad/ps was reported for CsPbBr$_3$ NCs in Ref.~\cite{Cai2023} and assigned to the exciton fine structure. We show below in Sec.~\ref{sec:fluct} that in our case the zero field oscillations originate from hole spin precession in the random hyperfine field of the nuclear spin fluctuations.

Figure~\ref{fig:Magnetic}(b) shows the same TRFR dynamics as in Fig.~\ref{fig:Magnetic}(a), but at larger time delays up to 5~ns, using an enlarged vertical scale. One identifies here the presence of a slowly oscillating component in the spin dynamics. Namely, the minimum of the signal is observed at about 3.5~ns for $B=45$~mT and shifts to smaller delays with increasing magnetic field. The amplitude of this slow component decreases with growing magnetic field. A fit with Eq.~\eqref{eq:VoightTRFR} gives the Larmor precession frequency $\omega_\text{L,slow}$, which magnetic field dependence is presented in Fig.~\ref{fig:Magnetic}(d). For $B<100$~mT, $\omega_\text{L,slow}$ depends linearly on magnetic field with $|g_\text{slow}|=0.09$. For larger magnetic fields $\omega_\text{L,slow}$ saturates and does not depend on magnetic field. Note that a small $g$-factor of $0.07$ was reported for CsPb(Cl,Br)$_3$  NCs (sample \#3), measured by optically detected magnetic resonance. It was attributed to spatially indirect excitons in the NCs, where one of the carriers is localized in the vicinity of the NC surface~\cite{Belykh2022}. For this sample, the decrease of the FR signal amplitude and the saturation of the Larmor precession frequency in magnetic field also were observed.

The dependence of the spin properties on NC size can be obtained by studying samples with differently sized NCs and/or studying the spin dynamics for one sample at different laser energies, which can be related to different NC sizes. We use both approaches for the CsPbBr$_3$ NCs. For measuring spectral dependences, the time-resolved Faraday ellipticity (TRFE) technique is more favorable than TRFR~\cite{Yugova2009}. Both techniques give the same information on $g$-factors and spin dephasing times, but for TRFR the dependence has a derivative-like shape approaching zero amplitude around the optical resonance, while TRFE does not vanish there. As an example, one can see the spectral dependences of the FE (red symbols) and FR (green symbols) amplitudes measured in the sample \#1 at $T=300$~K, see Fig.~\ref{fig:Temperature}(d). This sample has a broad size dispersion and by tuning the laser energy to larger values we address NCs with smaller sizes.

The spectral dependences of the initial FE amplitude at zero time delay, which is proportional to the induced spin polarization $S_0$, are shown in Figure~\ref{fig:Spectral}(a). Typically, these dependences correspond to the exciton absorption profiles of the NCs. They correlate well with the PL spectra, but with a high energy shift which scales with the inhomogeneity of the NC size distribution being largest in sample \#1.  

Figure~\ref{fig:Spectral}(b) shows the  $g$-factor dependence on the laser photon energy $E_\text{L}$. For the CsPbBr$_3$ NCs (samples \#1 and \#2) the $g$-factor increases with increasing energy (decreasing NC size). For the largest measured NC size of about 16~nm in sample \#2, we obtain $g=0.85$ at $E_\text{L} = 2.36$~eV. This value is close to the hole $g$-factor in bulk CsPbBr$_3$ ($g_\text{h,bulk}=0.75$). Further, the measured $g$-factor in the CsPbBr$_3$ NCs increases with growing energy up to 2.55~eV, above which it saturates up to 2.61~eV. 

The dashed lines in Fig.~\ref{fig:Spectral}(b) show the universal dependences of the electron and hole $g$-factors on the band gap energy in bulk lead halide perovskites. They were determined experimentally and confirmed theoretically in Ref.~\cite{kirstein2022nc}. No such experimental dependences are available for CsPbBr$_3$ NC so far, while model calculations~\cite{nestoklon2023_nl}, shown in Fig.~\ref{fig:Spectral}(b) by the solid lines, predict that the electron $g$-factor should decrease steeply with decreasing NC size (increasing effective band gap), while the hole $g$-factor should increase slowly remaining close to the universal dependence for bulk. Our experimental data deviate strongly from the model predictions. 

Let us first identify what type of carrier is responsible for the experimentally measured spin dynamics. We conclude that it is the hole, as in large NCs the value is close to the hole $g$-factor in bulk CsPbBr$_3$ and the $g$-factor demonstrates an increasing trend with energy, which is expected for holes. However, decrease of the NC size has a stronger effect on the hole $g$-factor than predicted theoretically. The underlying mechanisms still need to be identified, which goes beyond the scope of the present study. We note that we recently found a very similar behavior for CsPbI$_3$ NCs in glass~\cite{Meliakov_paper2}, which indicates that the effect is general for perovskite NCs and not limited to a specific material system.

\subsection{Temperature dependence of  spin dynamics}

\begin{figure*}[hbt!]
\centering
\includegraphics[width=1.8\columnwidth]{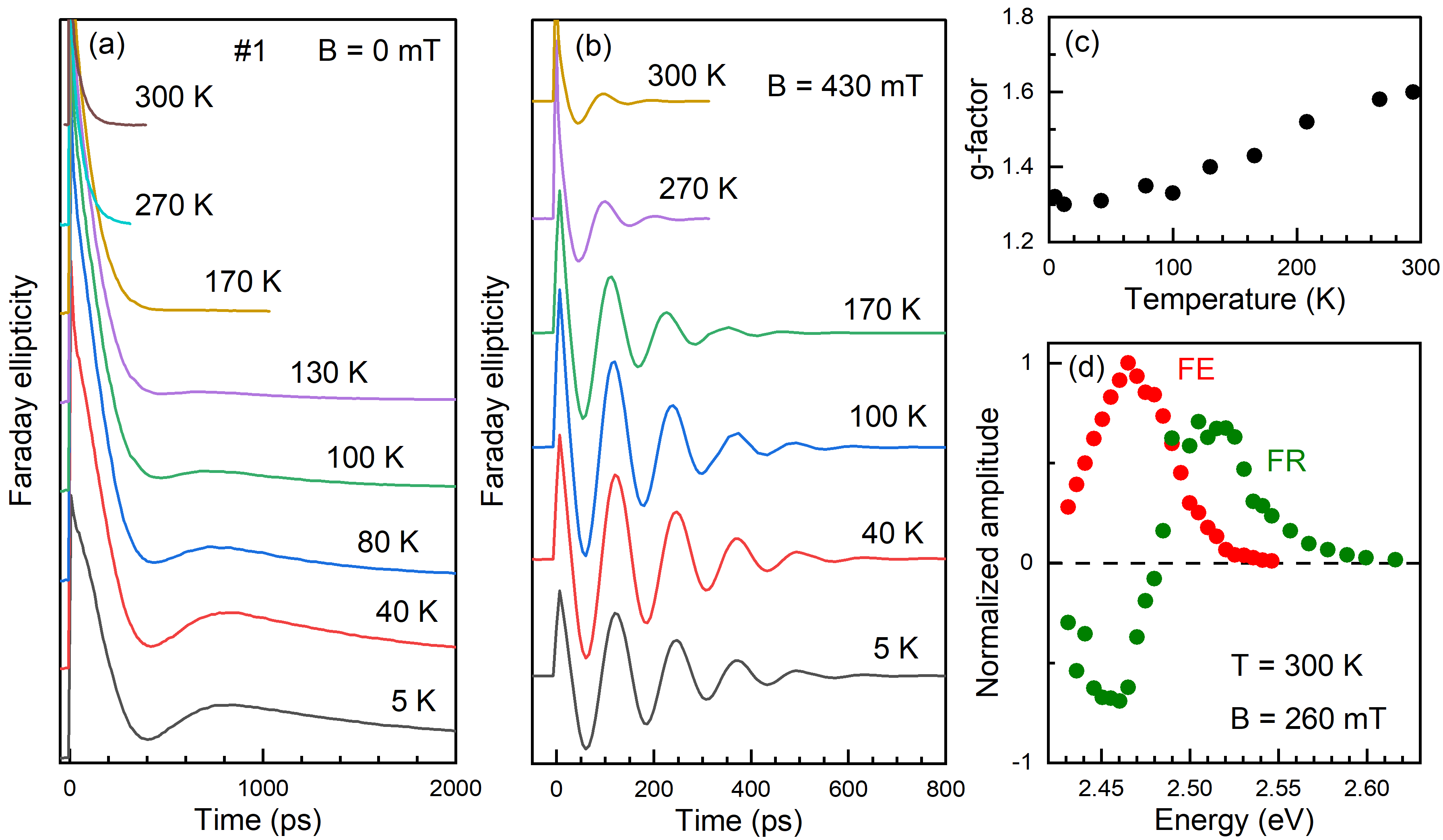} 
\caption{Temperature dependence of spin dynamics in CsPbBr$_3$ NCs (sample \#1).
(a) TRFE traces in zero magnetic field measured at temperatures varied from 5 to 300~K. 
(b) TRFE dynamics in the magnetic field of 430~mT measured at temperatures varied from 5 to 300~K.
(c) Temperature dependence of $g$-factor.
(d) Spectral dependence of FE (red symbols) and FR (green) amplitude at zero time delay measured at room temperature for $B=260$~mT.}
\label{fig:Temperature}
\end{figure*}

It is remarkable that in CsPbBr$_3$ NCs, coherent spin dynamics of carriers can be detected even at room temperature~\cite{Crane2020,Lin2022,Meliakov2023NCs,Cheng2024}. Here we demonstrate this finding for sample \#1, for which we can measure TRFE signals in the temperature range from 5 to 300~K, see Fig.~\ref{fig:Temperature}. For these measurements the laser photon energy in each case is adjusted to the maximum of the FE signal, which is $E_\text{L}=2.480$~eV at $T<$80~K and $2.475$~eV at higher temperatures. In the spin dynamics at zero magnetic field, shown in Fig.~\ref{fig:Temperature}(a), spin precession is observed from the lowest temperature up to 130~K. The frequency of these oscillations stays approximately constant at about 7~rad/ns.  At higher temperatures the spin dynamics demonstrate a monotonic decay. 

Figure~\ref{fig:Temperature}(b) shows the FE dynamics at $B=430$~mT measured at various temperatures. Pronounced spin oscillations are seen in the whole temperature range. At $T=5$~K the spin dephasing time is about 400~ps. It stays constant up to $T=100$~K and decreases at higher temperatures reaching 50~ps at room temperature.

\begin{figure}[hbt!]
\includegraphics[width=1\columnwidth]{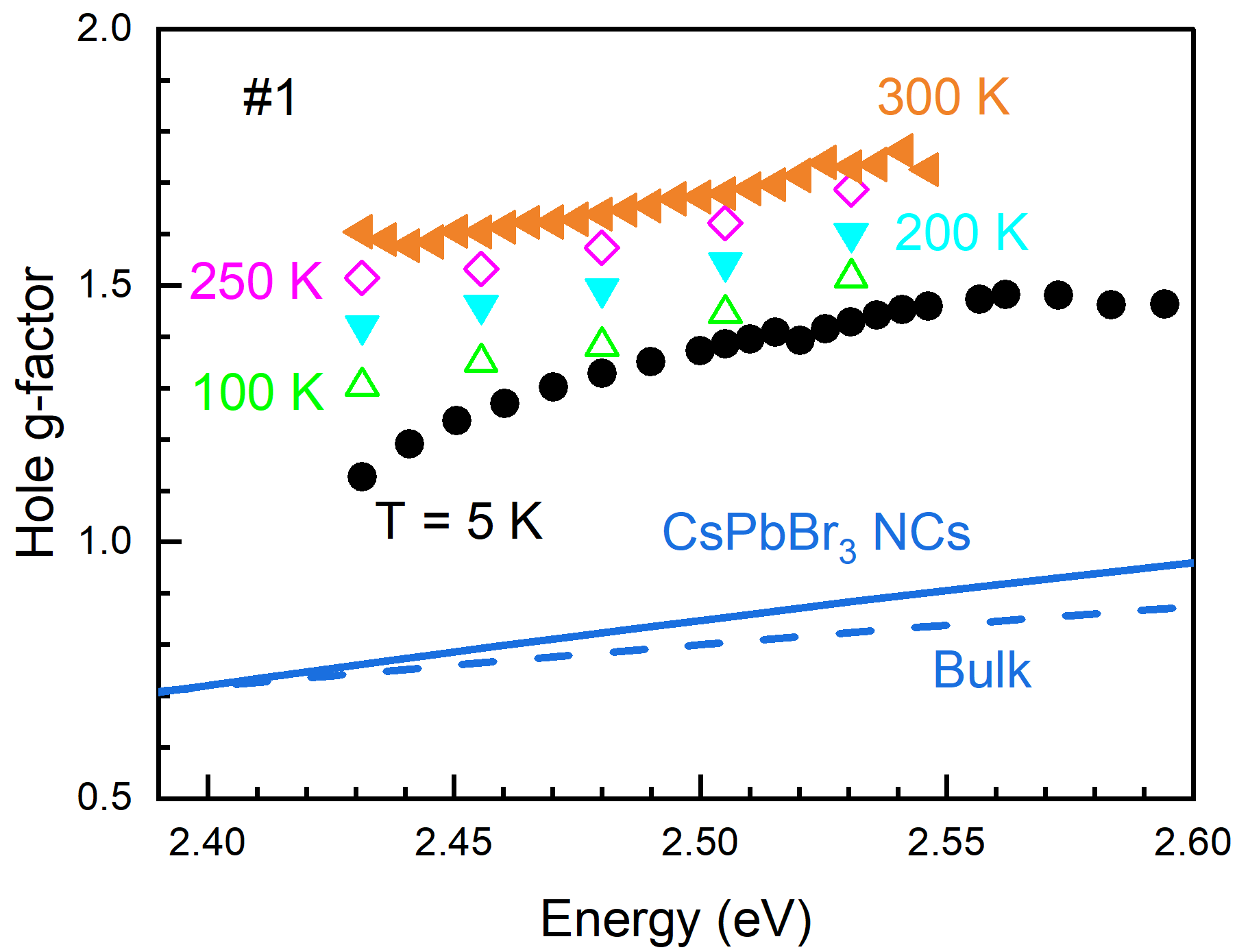} 
\caption{Spectral dependences of $g$-factor in CsPbBr$_3$ NCs (sample \#1) at temperatures of 5~K (black symbols), 100~K (green symbols), 200~K (cyan symbols), 250~K (pink symbols) and 300~K (orange symbols). The dashed line corresponds to the universal dependence of hole $g$-factor on the band gap energy in lead halide perovskites from Ref.~\onlinecite{kirstein2022nc}. The solid line shows the dependence of the hole $g$-factor on the band gap (i.e., the NC size), calculated for CsPbBr$_3$ NCs in Ref.~\onlinecite{nestoklon2023_nl}.
}
\label{fig:gTemp}
\end{figure}

Figure~\ref{fig:Temperature}(c) shows that the hole $g$-factor increases with temperature from 1.30 at $5$~K to 1.60 at room temperature. In order to examine whether such change can be related to the temperature variation of the band gap energy or whether other mechanisms are involved, we measure the spectral dependence of the $g$-factor at different temperatures, see Fig.~\ref{fig:gTemp}. The dependences at 100~K, 200~K and 250~K are taken the same sample spot, while the data for 5~K and 300~K are measured at different points. Obviously the $g$-factor increases with growing temperature in the whole spectral range. But the spectral range itself, across which spin dynamics are observed, does not significantly change with temperature. From that we conclude that in CsPbBr$_3$ NCs the temperature dependence of the hole $g$-factor is not determined by the temperature shift of the band gap, but other mechanisms which renormalize the involved band states. Disclosing these mechanisms, which would allow a deeper understanding of the electronic band structure of the lead halide perovskites and its modification in NCs, is a challenging task for theory and experiment.  

\begin{figure*}[hbt!]
\centering
\includegraphics[width=1.5\columnwidth]{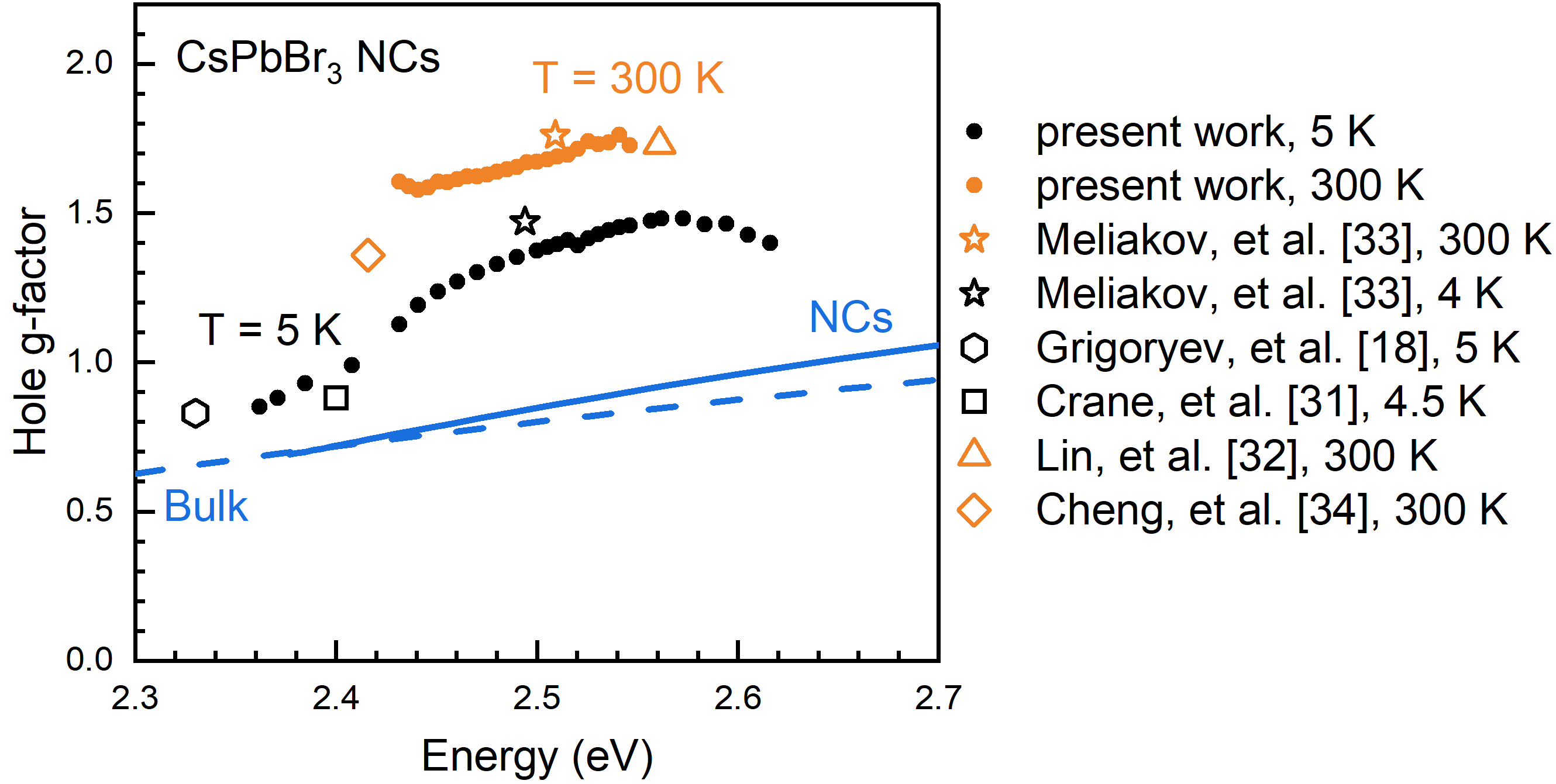}
\caption{Values of the hole $g$-factor in CsPbBr$_3$ NCs at cryogenic temperature (black symbols) and room temperature (orange symbols). Closed circles correspond to the values for CsPbBr$_3$ NCs in a glass matrix from this report. Open symbols show the data for colloidal NCs from Refs.~\onlinecite{Meliakov2023NCs,Grigoryev2021,Crane2020,Lin2022,Cheng2024}.
}
\label{fig:5-comp}
\end{figure*}

In Figure~\ref{fig:5-comp} we compare the values of the hole $g$-factor obtained in the present work with the data available for CsPbBr$_3$ colloidal NCs from Refs.~\onlinecite{Meliakov2023NCs,Grigoryev2021,Crane2020,Lin2022,Cheng2024}. The data are shown for $T \approx 5$~K and room temperature. One can see that the data for colloidal NCs and NCs in a glass matrix are in good agreement with each other. Note that in Ref.~\onlinecite{Meliakov2023NCs}, the $g$-factors of 1.47 at $T=5$~K and of 1.76 at room temperature were assigned to the electron, since, according to the theoretical predictions of Ref.~\onlinecite{nestoklon2023_nl}, the hole $g$-factor should be almost independent of NC size. In the present work we revise this interpretation, as we find the hole $g$-factor, starting almost from the ``bulk'' value, to be strongly dependent on the NC size (the transition energy). In what follows we confirm the ``hole assignment'' by measuring a strong hyperfine interaction with the nuclei spins as expected for holes rather than for electrons in perovskites. 

\subsection{Coherent spin dynamics in CsPb(Cl,Br)$_3$ NCs}

\begin{figure*}[hbt!]
\centering
\includegraphics[width=1.7\columnwidth]{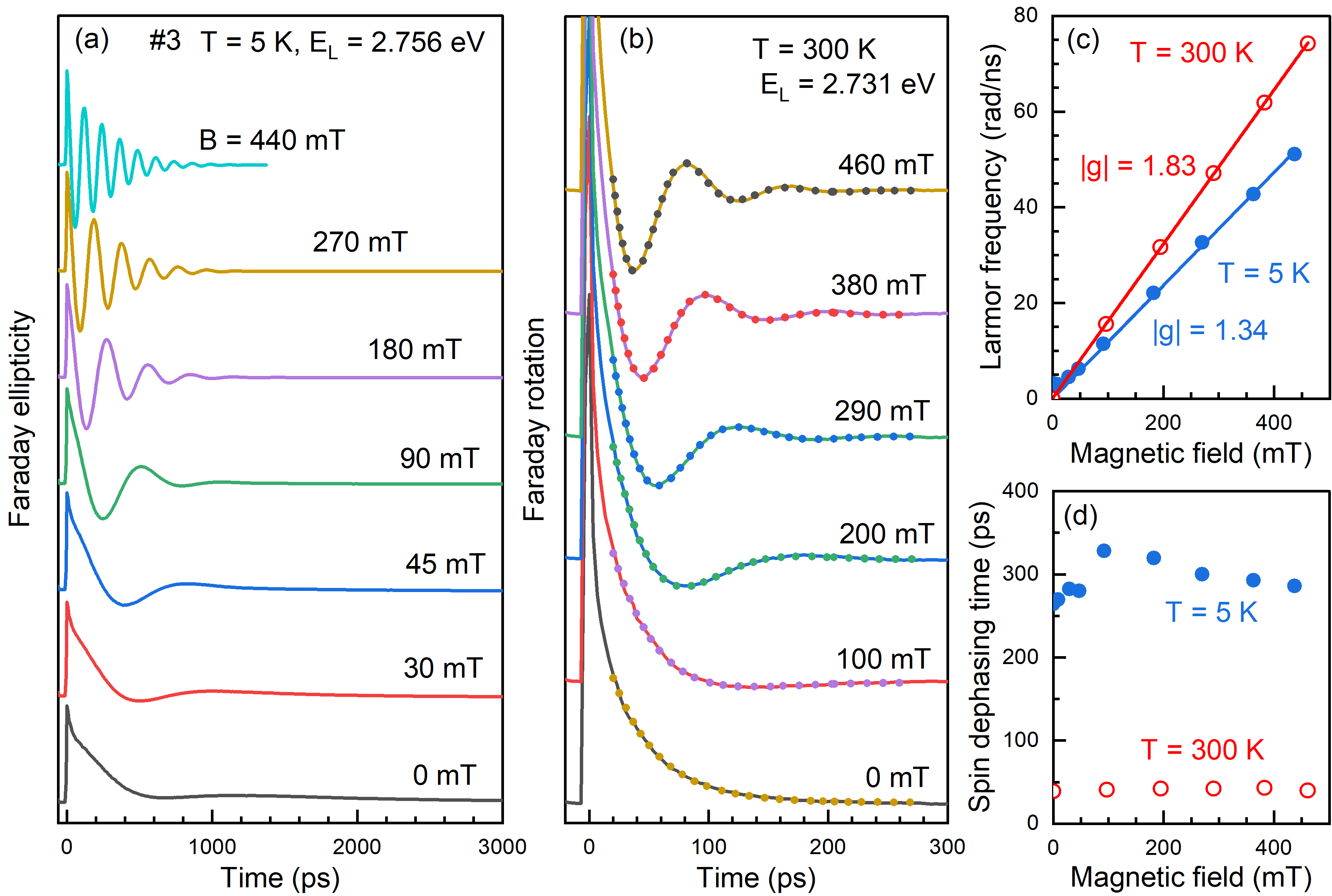}
\caption{Magnetic field dependence of the spin dynamics in CsPb(Cl,Br)$_3$ NCs (sample \#3).  
(a) FE dynamics measured at $T=5$~K in various magnetic fields. $E_\text{L}=2.756$~eV. 
(b) FR dynamics measured at $T=300$~K in various magnetic fields. $E_\text{L}=2.731$~eV. The experimental dynamics are shown by lines and their fits with Eq.~\eqref{eq:VoightTRFR} by symbols.
(c) Magnetic field dependence of the Larmor precession frequency at $T=5$~K (red symbols) and 300~K (blue symbols).  Lines are fits to the data with Eq.~\eqref{eq:LarmFreq}.
(d) Dependence of the spin dephasing time $T_2^*$ on magnetic field at $T=5$ and 300~K. 
}
\label{fig:SI-EK14LowT}
\end{figure*}

The phenomenology of the spin dynamics in CsPb(Cl,Br)$_3$ NCs is similar to that of the dynamics in CsPbBr$_3$ NCs. It is summarized in Fig.~\ref{fig:SI-EK14LowT}, where representative data of sample \#3 are shown for $T=5$~K and room temperature. At low temperature, one observes spin precession even at zero magnetic field. Its frequency is to about 3~rad/ns corresponding to the Zeeman spliting of 2~$\mu$eV. With increasing magnetic field, pronounced spin precession with a Larmor frequency corresponding to $|g|=1.34$ is seen. The magnetic field dependence of the Larmor precession frequency is given in Fig.~\ref{fig:SI-EK14LowT}(c) by the blue color symbols. The spin dephasing time $T_2^*$ of about 300~ps weakly depends on magnetic field, see Fig.~\ref{fig:SI-EK14LowT}(d).

The sample \#3 has a small spread of NC sizes and here the spin dynamics can be measured only in the narrow spectral range from 2.713 to 2.768~eV, see Fig.~\ref{fig:Spectral}(a). The measured values of the carrier $g$-factor of about 1.34 are close to the electron $g$-factor in bulk, see Fig.~\ref{fig:Spectral}(b). One should be careful with conclusions from this information, as even in bulk perovskites [dashed lines in Fig.~\ref{fig:Spectral}(b)] at the corresponding energy the electron and hole $g$-factors are not far from each other, and the confinement in NCs can bring them even closer. We have studied the same sample \#3 in Ref.~\onlinecite{kirstein2023_SML}, measured the $g$-factor of $1.2$ at $E_\text{L} = 2.74$~eV and assigned it to the hole. This assignment was based on observation of a strong dynamic nuclear polarization effect, which in lead halide perovskites is much stronger for the holes than for the electrons. It is worthwhile to note that in Fig.~\ref{fig:Spectral}(b) one should not consider the spectral dependence for the CsPb(Cl,Br)$_3$ NCs as an extension of that for the CsPbBr$_3$ NCs. In CsPb(Cl,Br)$_3$ NCs the optical transition energy is dominated by the band gap increase due to the Cl inclusion, and not so much by the confinement-induced shift. These two energy shifts contribute differently to the modification of the carrier $g$-factors.

The spin dynamics in CsPb(Cl,Br)$_3$ NCs measured at room temperature are shown in Fig.~\ref{fig:SI-EK14LowT}(b).
The spin dephasing time shortens down to the value $T_2^* \approx 40$~ps, independent of the applied magnetic field [Fig.~\ref{fig:SI-EK14LowT}(d)], but spin precession is clearly visible at magnetic fields exceeding 250~mT. The slope of the Larmor frequency dependence on the magnetic field (Fig.~\ref{fig:SI-EK14LowT}(c)) gives $|g|=1.83$. Thus, the $g$-factor grows from 1.29 to 1.83 with the temperature increasing from 5 to 300~K in this sample. 

\clearpage

\subsection{Role of nuclear spin fluctuations in the coherent spin dynamics of holes}
\label{sec:fluct}

In this section we discuss the phenomena in the coherent spin dynamics of the holes in perovskite NCs, which are provided by the interaction of their spins with the hyperfine fields of the nuclear spin fluctuations. We apply here the theory developed by Merkulov, Efros, and Rosen in Ref.~\onlinecite{Merkulov2002}, which considers the electron spin dephasing by nuclear spin fluctuations in GaAs semiconductor quantum dots (QDs). We refer to it as the MER model and describe its details in the Appendix. In contrast to III-V semiconductors, where the hyperfine interaction with the nuclear spins is strong for the electrons, in lead halide perovskites it is strong for the valence band holes. As the holes in perovskite semiconductors in the vicinity of the band gap show a simple band structure with spin $1/2$, the MER model can be used without revision.

The influence of nuclear spin fluctuation may be identified experimentally by (at least) two signatures in the coherent spin dynamics of carriers. The first one is spin precession at zero external magnetic field, due to carrier spin splitting in the hyperfine nuclear field. This precession can be masked due to its long period (compared to spin relaxation by other mechanisms or carrier lifetime).  However, we can detect it clearly in the studied CsPbBr$_3$ and CsPb(Cl,Br)$_3$ NCs, as shown in Figs. \ref{fig:Magnetic}(a), \ref{fig:Temperature}(a), and \ref{fig:SI-EK14LowT}(a). The second signature may be found in the spin precession decay, namely a nonexponential decay, when one deals with spin dephasing of an NC ensemble. 

According to the MER model, a carrier spin in a single NC precesses about the effective nuclear hyperfine magnetic field which is provided by the nuclear spin fluctuation. For simplicity, we refer to it as the hyperfine nuclear field or the nuclear field. This field has random orientation and amplitude, which change in time. But the characteristic time of their changes is much longer than the carrier spin precession period, so that one can consider the nuclear fields as ``frozen''. In order to model the spin dynamics in the NC ensemble, averaging over different directions and magnitudes of the hyperfine nuclear field needs to be done. According to the MER model, we take a Gaussian probability density distribution for the nuclear fields [Eq.~\eqref{eq:S_distr}], with $\Delta_B$ being the dispersion of the nuclear field distribution.     

In zero magnetic field the dynamics of the average spin projection of the carrier spin ensemble on the direction of their initial orientation (taken as $z$-axis, parallel to the pump beam) is described by
\begin{multline}
\left<S_z (t)\right> = \\ \frac{S_0}{3} \left\{ 1+2 \left( 1-2 \left( \frac{t}{2T_\Delta} \right) ^2 \right) \exp \left[ - \left( \frac{t}{2T_\Delta} \right)^2 \right] \right\}.
\label{eq:MERnoField}
\end{multline}
This corresponds to Eq.~(10) from Ref.~\onlinecite{Merkulov2002}, where $T_\Delta$ is corrected for $2T_\Delta$ \cite{Liang2017}. Here $T_\Delta$ is the effective spin dephasing time of the ensemble of carrier spins 
\begin{equation}
T_\Delta = \frac{\hbar}{|g| \mu_B \Delta_B} = \frac{\hbar}{\Delta_E}.
\label{eq:TDelta}
\end{equation}
$\Delta_E=|g| \mu_B \Delta_B$ is the dispersion of the carrier spin splitting in the nuclear field. According to Eq.~\eqref{eq:MERnoField}, the carrier spin polarization first approaches its minimal value at $t = \sqrt{6} T_\Delta$ and then recovers to a constant value of $S_0 / 3$. This variation can be described as half-period spin precession with a frequency that corresponds to about the dispersion $\Delta_E$: $\omega_L(B=0) \approx \Delta_E/\hbar$. In transverse external magnetic fields $B \gg \Delta_B$, the MER model gives a Gaussian decay of the spin precession with the characteristic time $\sqrt{2} T_\Delta$, see Eq.~\eqref{eq:S_MERmodel} in the Appendix. This time is about equal to the spin dephasing time received by exponential fit of the spin dynamics with Eq.~\eqref{eq:VoightTRFR}: $T_2^* \approx \sqrt{2} T_\Delta$. 

In order to account for the spin relaxation mechanisms not related to the nuclear fluctuations, e.g., responsible for decay of the $S_0 / 3$ polarization at longer delays, we add an exponential decay factor $\exp(-t/\tau_s^*)$ in the expression for $\left<S_z (t)\right>$, see Eqs.~\eqref{eq:S_MERmodel} and \eqref{eq:S_MERnoField}. Here $\tau_s^*$ is the characteristic spin relaxation time provided by these mechanisms. 

\begin{figure*}[hbt!]
\centering
\includegraphics[width=1.7\columnwidth]{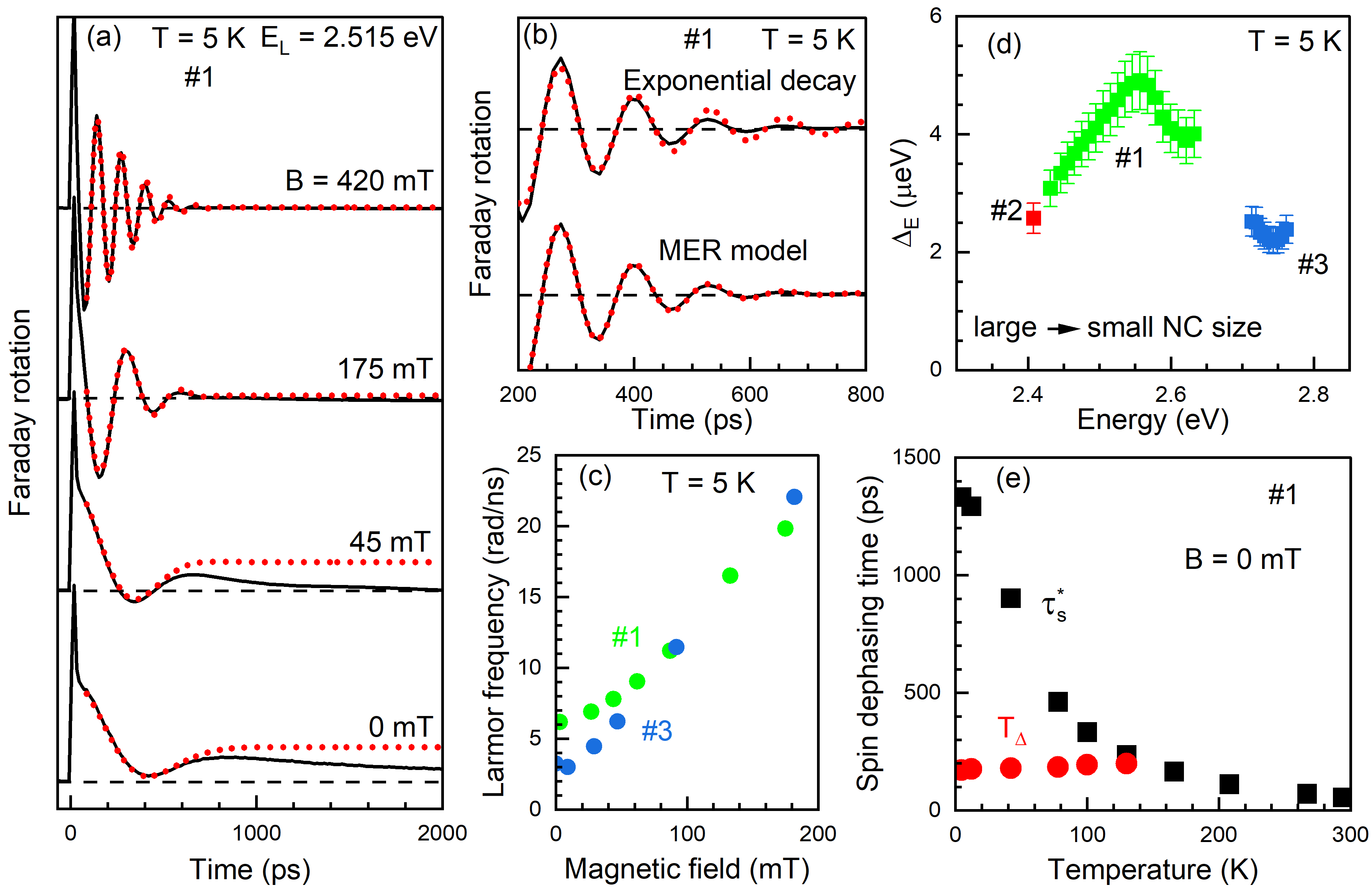} 
\caption{Hole spin dynamics in the hyperfine nuclear field of CsPbBr$_3$ and CsPb(Cl,Br)$_3$ NCs.
(a) TRFR dynamics measured on sample \#1 at various magnetic fields (solid lines). The dotted lines are fits with the MER model.
(b) Comparison of the fits to the TRFR signal at $B=430$~mT using Eq.~\eqref{eq:VoightTRFR}, accounting for a purely exponential decay (upper graph), and using Eq.~\eqref{eq:S_MERnoField} of the MER model with an exponential decay (lower graph).
(c) Magnetic field dependence of the Larmor precession frequency evaluated by fitting the spin dynamics with Eq.~\eqref{eq:VoightTRFR} for samples \#1 (green circles) and \#3 (blue circles).
(d) Spectral dependence of the dispersion $\Delta_E$ of the carrier spin splitting in the nuclear field. 
(e) Temperature dependence of the spin dephasing time $T_\Delta$ (red circles) and spin relaxation time $\tau_s^*$ (black squares) measured at zero magnetic field.
}
\label{fig:Fluct}
\end{figure*}

Let us turn to the analysis of the experimentally measured spin dynamics in CsPbBr$_3$ NCs (sample \#1) at $T=5$~K, presented in Fig.~\ref{fig:Magnetic}(a). We reproduce some of these dependences in Fig.~\ref{fig:Fluct}(a). As noted above, the observed spin precession at zero magnetic field cannot be described by Eq.~\eqref{eq:VoightTRFR}, which accounts only for the Zeeman splitting in external magnetic field. In our experiment at $B=0$~mT, the FR amplitude drops to almost zero for a delay time of about $400$~ps and then recovers to approximately 20\% of the initial value at a delay of about $900$~ps. The dotted line in the lower graph of Fig.~\ref{fig:Fluct}(a) shows a fit with the MER model Eq.~\eqref{eq:MERnoField}, which does not account for other mechanisms of spin relaxation described by $\tau_s^*$. The extracted effective spin dephasing time $T_\Delta = 170$~ps. The fit is in good agreement with the experiment at delays shorter than 800~ps, but does not account for the signal decay at longer delays. One can see that the calculated spin amplitude at long delays decreases from $S_0 / 3$ to almost zero as the magnetic field increases from 0 to $B>175$~mT, i.e., to $B \gg \Delta_B$. We note that fitting of the dynamics in a magnetic field $B>175$~mT at $T=5$~K does not require involvement of $\tau_s^*$ processes. Therefore, we determine the time $\tau_s^*$ by fitting the spin dynamics in zero magnetic field with  Eq.~\eqref{eq:S_MERnoField}, and then keep $\tau_s^*$ constant for higher fields. The respective fit with the evaluated parameters $T_\Delta = 170$~ps and $\tau_s^*=1300$~ps is shown in the lower graph of Fig.~\ref{fig:Magnetic}(a).   

In finite magnetic fields, the contribution of the nuclear fields to the decay of spin dynamics is clearly seen for sample \#1. In Figure~\ref{fig:Fluct}(b) we show the experimental dynamics measured at $B=430$~mT by black lines. Its decay cannot be described by the exponential function Eq.~\eqref{eq:VoightTRFR}, see the red dotted line in the upper graph. However, using the MER model approach one can closely fit the data using Eq.~\eqref{eq:S_MERmodel}, see the red dotted line in the lower graph. We use the MER model with the additional exponential decay (Eq.~\eqref{eq:S_MERmodel}) to fit the spin dynamics in various magnetic fields, see the dotted line in Fig.~\ref{fig:Magnetic}(a). It results in $T_\Delta = 170$~ps and $\tau_s^*=1300$~ps, which are independent of magnetic field strength. This value of $T_\Delta$ corresponds to $\Delta_E=4$~$\mu$eV and $\Delta_B=50$~mT, calculated with $g=1.30$. The fact that the decay of the spin precession at high magnetic fields is well-described by the time $T_\Delta$, determined from the fit of zero-field oscillations, confirms the consistency of our interpretation.

We obtain a good fit to the experimental data with $\tau_s^*$ independent of the magnetic field. However, we expect a decrease of $\tau_s^*$ in higher fields, where dephasing induced by the spread of $g$ factors, $\Delta g$, becomes relevant: $1/\tau_s^*(B) = 1/\tau_s^*(0) + \Delta g \mu_\text{B} B / \hbar$. Thus, at low magnetic fields $B \ll \Delta_B$ the redundant spin polarization decay is determined by the time $\tau_s^*(0)$. At the same time, the spin precession decays with the time $T_\Delta$ at the field $B \ll \hbar /|g| \mu_\text{B} T_\Delta$. At high magnetic fields, $B \gg \hbar /|g| \mu_\text{B} T_\Delta$, the spin precession decay is described by the time $\tau_s^*(B) = \hbar / \Delta g \mu_\text{B} B$ which decreases with $B$. The corresponding characteristic value of the magnetic field can be evaluated for CsPb(Br,Cl)$_3$ NCs in sample \#3 where $\Delta g=0.03$ was measured in \cite{kirstein2023_SML}. In this sample the spread of $g$-factors determines the spin precession decay for $B \gg 2$~T.

Figure~\ref{fig:Fluct}(c) shows a close-up of the magnetic field dependences of the signal oscillation frequency for samples \#1 and \#3. It illustrates the transition from the nuclear-induced spin precession at low external magnetic field to the well-established spin precession about the external field, corresponding to the linear region of the field dependence. We also clearly see that for sample \#1 the effective nuclear field is about twice as large as for sample \#3.

Figure~\ref{fig:Fluct}(d) shows the dependence of $\Delta_E$ on the laser photon energy. $\Delta_E$ is evaluated using Eq.~\eqref{eq:TDelta} from the times $T_\Delta$, which are obtained from the spin dynamics in zero magnetic field.  Here we collect the data for all studied samples. In  sample \#2, the zero-field spin precession is observed only at the energy of 2.41~eV and corresponds to  $\Delta_E \approx 2.6$~$\mu$eV. At smaller energies (larger NC sizes) a monotonic decay of the spin dynamics is observed at $B=0$~mT. In sample \#1, $\Delta_E$ increases with growing energy (decreasing NC size) from 3~$\mu$eV at $2.43$~eV to 5~$\mu$eV at $2.56$~eV. This behavior is in agreement with the expected dependence of $\Delta_E$ on the NC volume  (i.e. carrier localization volume $V_L$): ${\Delta_E}^2 \sim 1/V_L$~\cite{Petrov2008}. A further energy increase results in a small decrease of $\Delta_E$ to approximately 4~$\mu$eV. In the CsPb(Cl,Br)$_3$ NCs (sample \#3), the obtained value of $\Delta_E$ is about 2~$\mu$eV. We note that the average NC size in sample \#3 (8~nm) is close the NC size in sample \#1 (9~nm) and considerably smaller than the one in sample \#2 (16~nm). Thus, the smaller $\Delta_E$ in sample \#3 is presumably related to the introduction of Cl.

The measured Zeeman splitting in the hyperfine field corresponds to the effective nuclear field $\Delta_B$ ranging from 25 to 65~mT for samples \#1, \#2, and \#3. These relatively large nuclear fields are expected for the hyperfine interaction of the holes rather than the electrons in the perovskites \cite{kirstein2022am}, again confirming the hole nature of the observed spin precession.

Fitting the spin dynamics from Figs.~\ref{fig:Temperature}(a) at different temperatures with Eq.~\eqref{eq:S_MERnoField} allows us to evaluate the temperature dependences of the times $T_\Delta$ and $\tau_s^*$, see Fig.~\ref{fig:Fluct}(e). At the temperature of 5~K, $T_\Delta \ll \tau_s^*$. Therefore, the hole spin dephasing is controlled by the interaction with the hyperfine nuclear field. As the temperature increases, $T_\Delta$ stays about constant ($T_\Delta \approx 170$~ps) as long as zero field oscillations are observed. However, other mechanisms of spin relaxation accounted for by the time $\tau_s^*$ accelerate with temperature increase. At temperatures $T>130$~K, the time $\tau_s^*$ becomes shorter than $T_\Delta$, and the decay of the spin dynamics becomes monotonic. 
At room temperature the spin dynamics is determined only by the time $\tau_s^* \approx 50$~ps.

\begin{figure}[hbt!]
\includegraphics[width=1\columnwidth]{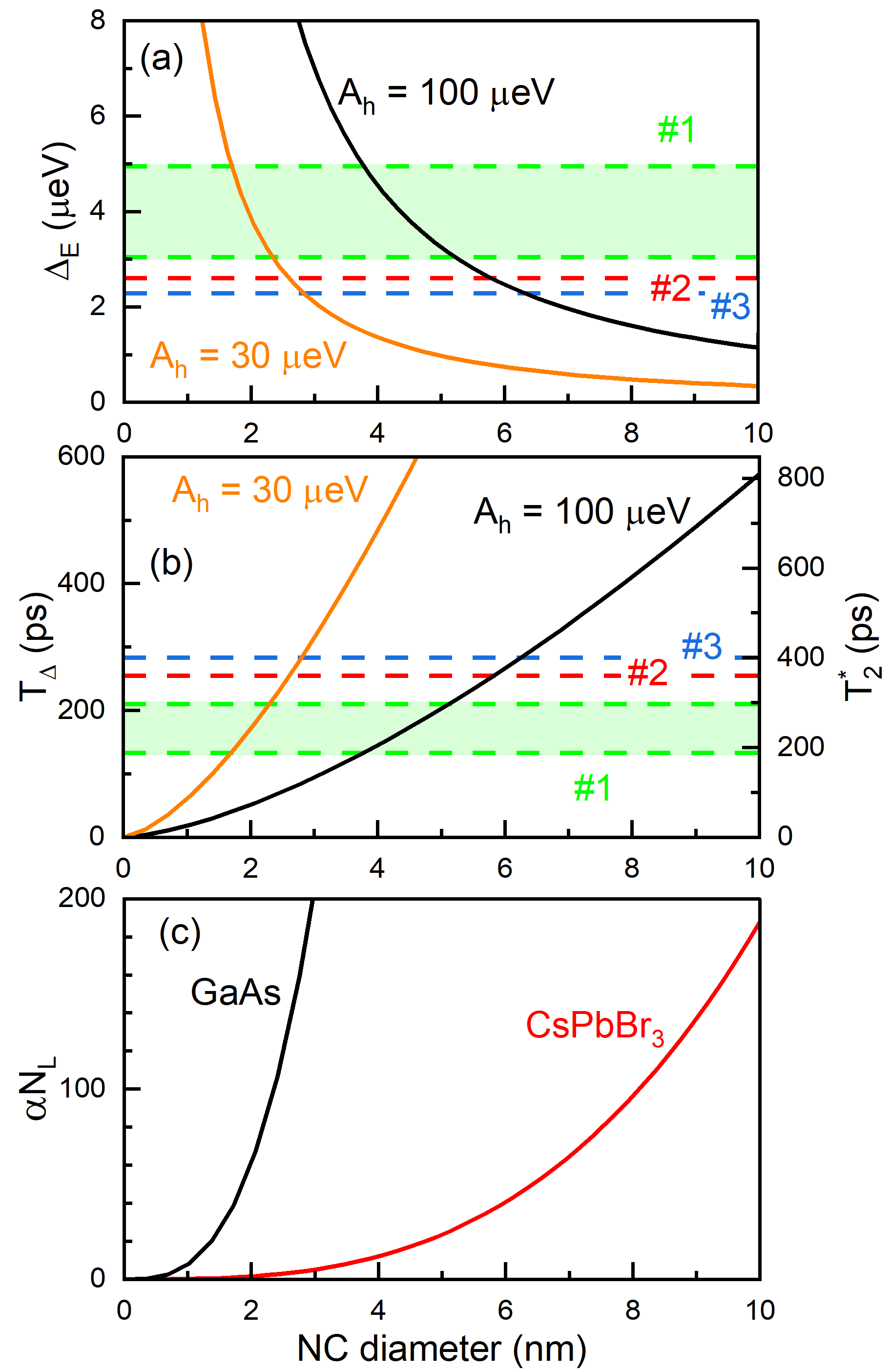} 
\caption{Evaluation of parameters responsible for the spin dynamics of a hole interacting with lead spins in perovskite spherical NCs. (a) Dependence of the dispersion $\Delta_E$ on NC diameter. (b) Dependence of the effective spin dephasing time $T_\Delta$ and the spin dephasing time $T_2^*=\sqrt{2}T_\Delta$ on NC diameter.
In panels (a,b) solid lines correspond to Eq.~\eqref{eq:S_HyperfineA1} with $A_\text{h} = 30$~$\mu$eV (orange line) and $A_\text{h} = 100$~$\mu$eV (black line). Green areas show range of experimental data for sample \#1. Red and blue dashed lines correspond to the experimental values for samples \#2 and \#3, respectively.
(c) Dependence of a number of the nuclei spins, $\alpha N_L$, interacting with a carrier localized in a NC on NC diameter. For CsPbBr$_3$ (red line, $^{207}$Pb isotopes with $\alpha=0.22$) and GaAs (black line, $\alpha=1$) NCs.
}
\label{fig:theory}
\end{figure}

The MER theory allows one to evaluate the spin dynamics parameters for carriers interacting with nuclear spin fluctuations. They depend of the carrier localization volume, $V_{L}$, which depends on NC size $d$ (diameter in case of spherical NC), the hyperfine interaction constant $A$ of the carrier with the nuclei, the nuclear spin $I$ and abundance of nuclear isotopes with nonzero spins $\alpha$. In Appendix A2 we adopt the formalism of the MER theory for the hole interaction with $^{207}$Pb nuclei in spherical lead halide perovskite NCs. The model results for the NC diameter dependences of $\Delta_E$ and $T_\Delta$ are shown in Figs.~\ref{fig:theory}(a) and~\ref{fig:theory}(b), respectively. They are calculated with Eq.~\eqref{eq:S_HyperfineA1} and the following parameters: the primitive cell size $a_0=0.595$~nm, nuclear spin $I=1/2$, and $^{207}$Pb abundance $\alpha=0.22$. In Ref.~\onlinecite{kirstein2022am} the hole hyperfine intraction constant was estimated in the range of $30-100$~$\mu$eV. Therefore, we plot two dependences with $A_h=30$~$\mu$eV (orange line) and $100$~$\mu$eV (black line). One can see, that with increasing NC diameter $\Delta_E$ decreases and $T_\Delta$ increases. This indicates that the nuclei have less effect on the hole spin dynamics, while the total number of involved nuclei grows.Two factors with opposite trends are involved here. First, the nuclear field fluctuation, is proportional to the square root of the number of nuclei in the NC volume, which is proportional to $d^{3/2}$. Second,  the contribution of the individual nuclei to the hyperfine field is proportional to the squared wave function amplitude of the carrier on it, which is inversely proportional to the NC volume ($\propto 1/d^{3}$).  As a result, the fluctuation of the hyperfine nuclear field depends on NC size as $\propto 1/d^{3/2}$. 

In order to relate the model considerations  with our experimental results we show in Figs.~\ref{fig:theory}(a) and~\ref{fig:theory}(b) the ranges for the experimental data by dashed lines. From the crossing points of the dashed and solid lines the hole localization volume can be evaluated. Namely, it allows one to estimate for which  NC sizes the respective experimental parameters are expected to be realized. The characteristic range is $2-6$~nm, which is smaller than the NC sizes of $7-16$~nm obtained by the sample characterization. Thus, we conclude that in the studied NCs holes are localized in smaller volumes, than expected from the NC diameter.

It is also instructive to know the  number of the nuclear spins interacting with a hole in a NC of specific diameter. We plot such dependence in Fig.~\ref{fig:theory}(c), which shows that for the studied perovskite NCs we are in range of $5-40$ nuclear spins. It is 40 times smaller than the number of nuclear spins in GaAs NC of the same size. For comparison, the evaluation for the GaAs NCs are shown by a black line, which is made for the unit cell of the size of  0.575~nm with 8 nuclei spins. Note that in GaAs all isotopes have spins. In this respect, the perovskite NCs offer a new model situation in spin physics of semiconductors when a confined carrier interacts with a small number of nuclear spins. In this regime the spin-flip of a single nuclei becomes  valuable, which should enrich spin dynamics of the carriers and nuclei spins.


\section{Conclusions}

To conclude, we have investigated the coherent spin dynamics of carriers using the pump-probe Faraday rotation/ellipticity technique in CsPbBr$_3$ and CsPb(Cl,Br)$_3$ perovskite nanocrystals. At cryogenic temperatures and in magnetic fields we observe oscillations of the signal with $g$-factors ranging from 0.8 to 1.5, depending on the NC optical transition energy. We attribute them to Larmor precession of the hole spins. A decrease of the NC size has stronger effect on the hole $g$-factor than predicted theoretically. The responsible mechanisms for this behavior  need to be identified in future. Coherent spin precession persists up to room temperature. The hole $g$-factor increases with temperature and the spin dephasing time shortens from about 1~ns at low temperatures to 50~ps at room temperature. Nuclear spin fluctuations play an important role in the hole spin dynamics via the hyperfine nuclear fields, which in the studied samples are in the range of 25 -- 65~mT. They modify the decay of the spin dynamics and also provide spin splitting of the holes at zero external magnetic field, which results in spin precession. This behavior follows the reference description presented in Ref.~\cite{Merkulov2002} and was so far observed only for a few material systems, but not for perovskite NCs. Unlike in other systems, we observe precession of the hole spin (rather than the electron spin) in the hyperfine field of the nuclear spin fluctuation, which is due to the specifics of the lead halide perovskite band structure compared to conventional semiconductors. Thereby, perovskite NCs show prominent spin properties, which can be easily tailored by NC size and composition, across a wide temperature range. This highlights perovskite semiconductors as promising candidates for spintronics applications.

\section*{Acknowledgments}
We acknowledge stimulating discussions with D. S. Smirnov, M. M. Glazov, K. V. Kavokin, and Al. L. Efros. Research performed at the P. N. Lebedev Physical Institute was financially supported by the Ministry of Science and Higher Education of the Russian Federation, Contract No. 075-15-2021-598.   E.V.K. and M.S.K. acknowledge support by the Saint-Petersburg State University (Grant No. 122040800257-5).

%

\textbf{AUTHOR INFORMATION}

{\bf Corresponding Authors} \\
Sergey~R.~Meliakov,  Email: melyakovs@lebedev.ru   \\
Dmitri R. Yakovlev,  Email: dmitri.yakovlev@tu-dortmund.de\\

\textbf{ORCID}\\
Sergey~R.~Meliakov         0000-0003-3277-9357 \\  
Vasilii~V.~Belykh:          0000-0002-0032-748X \\ 
Evgeny~A.~Zhukov:          0000-0003-0695-0093 \\  
Elena V. Kolobkova:        0000-0002-0134-8434 \\  
Maria S. Kuznetsova:       0000-0003-3836-1250 \\  
Manfred~Bayer:             0000-0002-0893-5949 \\ 
Dmitri R. Yakovlev:        0000-0001-7349-2745 \\  

\setcounter{equation}{0}
\setcounter{figure}{0}
\setcounter{table}{0}
\setcounter{section}{0}
\setcounter{subsection}{0}
\renewcommand{\theequation}{A\arabic{equation}}
\renewcommand{\thefigure}{A\arabic{figure}}
\renewcommand{\thetable}{A\arabic{table}}
\renewcommand{\thesubsection}{A\arabic{subsection}}

\begin{widetext}
\begin{center}

\section*{Appendix}

\end{center}

\subsection{Theory of spin dynamics of carriers interacting with nuclear spin fluctuations in NCs}
\label{Sec:SI_Merkulov}

The theoretical description of the electron spin dephasing by nuclear spin fluctuations in semiconductor quantum dots (QDs) was developed by Merkulov, Efros, and Rosen in Ref.~\onlinecite{Merkulov2002}. We refer to it as the MER model. The electron spin dynamics were considered for III-V semiconductor QDs, where the hyperfine interaction is known to be strong for the electron in conduction band with spin $1/2$, while the hole interaction with the nuclei is considerably weaker. In perovskite NCs both conduction band electrons and valence band holes have spin $1/2$, but the hyperfine interaction with the nuclei is considerably stronger for the holes~\cite{belykh2019,kirstein2022am,kudlacik2024_oo_carriers}. We present here the formalism that we used for modeling the spin dynamics in perovskite NCs. It is based on the MER model~\cite{Merkulov2002} and for the lead halide perovskites is universal for both holes and electrons.  

Following the MER model we consider the carrier spin precession about the magnetic field ${\bf B}_\text{S}$, which is the sum of the external magnetic field $\bf{B}$ and the effective nuclear hyperfine magnetic field $\textbf{B}_N$ in a single NC. We call, for simplicity, $\textbf{B}_N$ as the hyperfine nuclear field. One can introduce the unit vector ${\bf n}=\frac{{\bf B}+{\bf B}_N}{|{\bf B}+{\bf B}_N|}$, which points along the direction of $\textbf{B}_\text{S}$. Note that in zero external magnetic field it corresponds to the direction of the hyperfine nuclear field:  ${\bf n}=\frac{{\bf B}_N}{B_N}$. Thus, the spin precession in a single NC is described by the Bloch equation~\cite{Bloch1946,Abragam1961}:
\begin{equation}
\frac{d\mathbf{S}}{dt} = -\frac{g\mu_\text{B}}{\hbar} \mathbf{S} \times \textbf{B}_\text{S} - \frac{\mathbf{S} - \mathbf{S}_\text{st}}{\tau_s^*} \, .
\label{eq:Bloch}
\end{equation}
Here $\tau_s^*$ is the spin relaxation time accounting for the mechanisms of spin relaxation other than nuclei, and $\mathbf{S}_\text{st}$ is the stationary spin polarization in the field $\textbf{B}_S$. Taking into account that average initial spin polarization $\mathbf{S}_0$ in a NC is much grater than stationary spin polarization $\mathbf{S}_\text{st}$, the solution of this equation is given by 
\begin{equation}
\textbf{S}(t)=\{ (\textbf{S}_0 \cdot \textbf{n})\textbf{n} + [\textbf{S}_0-(\textbf{S}_0 \cdot \textbf{n})\textbf{n}]\cos{\Omega t} + [(\textbf{S}_0-(\textbf{S}_0 \cdot \textbf{n})\textbf{n})\times\textbf{n}] \sin{\Omega t} \}\exp(-t/\tau_s^*).
\label{eq:S_spin_prec}
\end{equation}
Here,  $\Omega = |g| {\mu_B |\textbf{B}+{\bf B}_N|/\hbar}$ is the Larmor precession frequency of the carrier, $g$ is the carrier $g$-factor, $\mu_B$ is the Bohr magneton, and $\hbar$ the reduced Planck constant. This equation corresponds to Eq.~(9) from Ref.~\onlinecite{Merkulov2002}, with the additional factor $\exp(-t/\tau_s^*)$ accounting for the spin relaxation processes, which are not related to the interaction with the nuclei. 

We analyze the dynamics of the spin polarization projection $S_z = \frac{(\textbf{S} \cdot \textbf{S}_0)}{S_0}$ onto the optical z-axis:
\begin{equation}
S_z(t) = \left\{S_0\cos{\Omega t} + \frac{(\textbf{S}_0 \cdot \textbf{n})^2}{S_0}(1-\cos{\Omega t})\right\} \exp(-t/\tau_s^*).
\label{eq:S_spin_proj}
\end{equation}
The external magnetic field $\textbf{B}$ is applied in the Voigt geometry (${\bf B} \perp z$). Therefore, the Eq.~\eqref{eq:S_spin_proj} can be rewritten as
\begin{equation}
S_z(t) = \left\{S_0\cos{\Omega t} + \frac{(\textbf{S}_0 \cdot {\bf B}_N)^2}{S_0|\textbf{B}+{\bf B}_N|^2}(1-\cos{\Omega t})\right\} \exp(-t/\tau_s^*).
\label{eq:S_spin_proj2}
\end{equation} 
Note that the carrier Larmor precesison in the hyperfine nuclear field ${\bf B}_N$ is much faster than the precession of a nucleus in the hyperfine field induced by a carrier (Knight field). Therefore, the carrier is subject to the "frozen fluctuation" of the nuclear field. In an ensemble of NCs, the vectors ${\bf B}_N$ are distributed according to a Gaussian probability density distribution~\cite{Merkulov2002}  
\begin{equation}
W({\bf B}_N) = \frac{1}{\pi^{3/2}\Delta_B^3} \exp\left( -\frac{|{\bf B}_N|^2}{\Delta_B^2} \right).
\label{eq:S_distr}
\end{equation}
Here, $\Delta_B$ is the dispersion of the nuclear hyperfine field distribution. Averaging of Eq.~\eqref{eq:S_spin_proj2} over the distribution Eq.~\eqref{eq:S_distr} gives us the dynamics of the projection of the spin polarization onto the optical axis in an ensemble of randomly oriented NCs:
\begin{equation}
\left<S_z(t)\right> = \frac{\exp({-t/\tau_s^*})}{\pi^{1/2}\Delta_B^3} \int\limits_{B_N=0}^{\infty}B_N^{2}dB_N \int\limits_{\theta=0}^{\pi}\sin{\theta}d\theta \left\{2 S_0\cos{\Omega t} + \frac{S_0 B_N^2 \sin^2{\theta}}{B^2+B_N^2+2{B}{B_N}\cos{\theta}}(1-\cos{\Omega t})\right\} \exp\left( -\frac{B_N^2}{\Delta_B^2} \right).
\label{eq:S_MERmodel0}
\end{equation}
Here, $\theta$ is the angle between the vectors $\textbf{B}$ and $\textbf{B}_N$. We can replace the integration parameter $B_N$ with $\omega_N = |g|\mu_{B}B_N$.

\begin{equation}
\left<S_z(t)\right> = \frac{\exp({-t/\tau_s^*}) T_\Delta^3}{\pi^{1/2}} \int\limits_{\omega_N=0}^{\infty}\omega_N^{2}d\omega_N \int\limits_{\theta=0}^{\pi} \sin{\theta} d\theta \left\{2 S_0\cos{\Omega t} + \frac{S_0 {\omega_N}^2 \sin^2{\theta}}{{\omega_\text{L}}^2+{\omega_N}^2+2{\omega_\text{L}}{\omega_N}\cos{\theta}}(1-\cos{\Omega t})\right\} \exp\left( -{\omega_N^2}{T_\Delta^2} \right).
\label{eq:S_MERmodel}
\end{equation}
Here, $\Omega$ can be presented in the form $\Omega^2 = \omega_\text{L}^2 + {\omega_N}^2 + 2{\omega_\text{L}}{\omega_N}\cos{\theta}$, where $\omega_\text{L} = |g| {\mu_B {B}/\hbar}$ is the Larmor precession frequency of the carrier about the external magnetic field in absence of the hyperfine nuclear field. The spin dephasing time $T_\Delta$ is determined according to Eq.~\eqref{eq:TDelta} from the main text. There we use Eq.~\eqref{eq:S_MERmodel} to fit the spin dynamics in an external magnetic field (the fit parameters are $T_\Delta$, $\omega_\text{L}$, $S_0$, and $\tau_s^*$). In zero external magnetic field, one obtains Eq.~\eqref{eq:MERnoField} with an additional exponential decay:


\begin{equation}
\left<S_z (t)\right> = \frac{S_0}{3} \left\{ 1+2 \left( 1-2 \left( \frac{t}{2T_\Delta} \right) ^2 \right) \exp \left[ - \left( \frac{t}{2T_\Delta} \right)^2 \right] \right\} \exp({-t/\tau_s^*}).
\label{eq:S_MERnoField}
\end{equation}
This equation differs from the one given in the MER model~\cite{Merkulov2002}. Namely, the arguments in the bracket are $t/{2T_\Delta}$ instead of $t/{T_\Delta}$ from Eq.~(10) in Ref.~\onlinecite{Merkulov2002}. Our version corresponds to Eq.~(2) of Ref.~\onlinecite{Liang2017}, where this error was noted. Also correct version of this equation can be found in Refs.~\onlinecite{GlazovBook2018} and~\onlinecite{Leppenen2022}.

\subsection{Spin dynamics of holes interacting with lead nuclei in perovskite NCs}

Let us evaluate the role of the lead spin fluctuations for the hole spin dynamics in the lead halide perovskite NCs. The lead has four different isotopes, but only one of them $^{207}$Pb has a non-zero spin ($I=1/2$) with corresponding abundance $\alpha=0.22$, i.e. 22\% of all lead ions have spin and contribute to hyperfine interaction with holes. There is one lead ion per unit cell composed of five ions, e.g. in CsPbBr$_3$ the unit cell for cubic lattice with lattice constant of $a_0 = 0.595$~nm  has one Pb, one Cs and three Br ions.

According to Eq.~(A7) from Ref.~\onlinecite{Syperek2011} we can evaluate the effective spin dephasing time $T_\Delta$ and, respectively, the dispersion of hyperfine interaction energy fluctuations $\Delta_E$ and the spin dephasing time $T_2^*$ as:
\begin{equation}
T_\Delta =  \frac{T_2^*}{\sqrt{2}} = \frac{\hbar}{\Delta_E}= \hbar \sqrt{\frac{3 N_\text{L}}{2 I(I+1)\alpha A_h^2}}  \,.
\label{eq:S_HyperfineA0}
\end{equation}
$N_L$ is a number of lead nuclei in the hole localization volume $V_L$, which is commonly smaller than the NC volume. In order to estimate $N_L$ one can use equation $N_L = V_L/a_0^3$, where $a_0$ is a size of the unit cell. In our case we can use primitive cell with single lead nucleus. $A_h$ is a constant of hyperfine interaction of hole and lead nucleus.
Using the wave function $\Psi$ for a hole in a spherical NC with diameter $d$ from Ref.~\onlinecite{ivchenko05a} (Eq.~(2.89)) one can estimate the volume of hole localization in a NC:
\begin{equation}
V_L^{-1} = \left(\int\Psi^4(\bm r) d\bm{r}\right)^{-1} = \int\limits_0^{d/2}\left(\frac{\sin(2\pi r/d)}{\sqrt{\pi d}r}\right)^44\pi r^2dr \approx \frac{5.4}{d^3} \,.
\label{eq:S_VL}
\end{equation}
Note, that in a NC with infinitely high barriers the hole localization volume $V_L=0.185d^3$ is only 36\% of the NC volume $V_{NC}=\pi d^3/6=0.52d^3$. Thus, Eq.~\eqref{eq:S_HyperfineA0} can be rewritten in the following form:
\begin{equation}
T_\Delta =\frac{T_2^*}{\sqrt{2}} = \frac{\hbar}{\Delta_E} \approx \hbar \sqrt{\frac{0.28d^3}{a_0^3 I(I+1)\alpha A_h^2}}\,.
\label{eq:S_HyperfineA1}
\end{equation}

\end{widetext}

\clearpage

\end{document}